\documentclass[11pt,amsmath,superscriptaddress,nofootinbib]{revtex4}

\usepackage[english]{babel}
\usepackage{epsfig}

\usepackage[utf8]{inputenc}
\usepackage{floatrow}
\usepackage{amssymb,amsmath,amsfonts,amsthm,graphicx,psfrag}
\usepackage{indentfirst}
\usepackage{hyperref}

\usepackage[title,titletoc]{appendix}
\usepackage{bm}
\usepackage{verbatim}
\usepackage{epsfig}

\usepackage{mathrsfs}
\usepackage[T1]{fontenc}

\usepackage{slashed}
\usepackage{enumitem}
\newcommand{\be}{\begin{eqnarray}}
\newcommand{\ee}{\end{eqnarray}}

\def\fun#1#2{\lower3.6pt\vbox{\baselineskip0pt\lineskip.9pt
\ialign{$\mathsurround=0pt#1\hfil ##\hfil$\crcr#2\crcr\sim\crcr}}}

\newcommand{\tr}{\operatorname{tr }}
\renewcommand{\Re}{\operatorname{Re }}
\renewcommand{\Im}{\operatorname{Im }}

\newcommand{\vep}{\mbox{\boldmath${\rm p}$}}

\newcommand{\lan}{\langle}
\newcommand{\ran}{\rangle}

\usepackage{color}

\begin{document}

\title{Dense Quark-Gluon Plasma in strong magnetic fields}

\author{R.A.Abramchuk}
\email{abramchuk@phystech.edu}
\affiliation{Moscow Institute of Physics and Technology, 9, Institutskii per., Dolgoprudny, Moscow Region, 141700, Russia}
\affiliation{Institute for Theoretical and Experimental Physics of NRC ``Kurchatov Institute'', B. Cheremushkinskaya 25, Moscow, 117259, Russia}
\author{M.A.Andreichikov}
\affiliation{Moscow Institute of Physics and Technology, 9, Institutskii per., Dolgoprudny, Moscow Region, 141700, Russia}
\affiliation{Institute for Theoretical and Experimental Physics of NRC ``Kurchatov Institute'', B. Cheremushkinskaya 25, Moscow, 117259, Russia}
\author{Z.V Khaidukov}
\affiliation{Moscow Institute of Physics and Technology, 9, Institutskii per., Dolgoprudny, Moscow Region, 141700, Russia}
\affiliation{Institute for Theoretical and Experimental Physics of NRC ``Kurchatov Institute'', B. Cheremushkinskaya 25, Moscow, 117259, Russia}
\author{Yu.A.Simonov}
\affiliation{Institute for Theoretical and Experimental Physics of NRC ``Kurchatov Institute'', B. Cheremushkinskaya 25, Moscow, 117259, Russia}

\begin{abstract}
    A non-perturbative (np) method of Field Correlators (FCM) was applied to study QCD at temperatures above the deconfinement transition (\(1<T/T_c<3,~T_c\sim0.16~GeV\)) and nonzero baryon densities (baryon chemical potential \(\mu_B<0.5~GeV\)) in an external uniform magnetic field (\(eB<0.5~GeV^2\)).
    Within FCM, the np high-temperature dynamics is embodied in the Polyakov loop and in the Debye mass due to   the Color-Magnetic Confinement.
    Analytic expressions for quark pressure and magnetic susceptibility were obtained.
    The expressions were represented as series and in integral form.
    Magnetic susceptibility was found to increase rapidly with temperature and slowly with density.
    The results at the zero density limit are in agreement with lattice data.

\end{abstract}
\maketitle

\section{Introduction}\label{sect_intro}

Strong magnetic fields emerge in various areas of physics, e.g.~in cosmology \cite{1,2}, in non-central heavy-ion collisions \cite{3,4,5,6}, in neutron stars physics \cite{7}; see \cite{8} for review on strongly interacting matter in magnetic fields.  

The influence of a magnetic field (m.f.) on QCD thermodynamics was studied in many model approaches \cite{9}. 
In particular, for the results within the Nambu--Jona--Lasinio (NJL) model and the holographic approach see \cite{10,11} and \cite{12}, respectively.

The principal problem of a strongly interacting system description is its non-perturbative dynamics.
In the present paper, we address the problem with the Field Correlator Method \cite{13} (see \cite{16} for a recent review).
The QCD thermodynamics within FCM was initially developed in \cite{17} with incorporation of Polyakov loops (for a comparison with lattice data, see \cite{18}).

Recently, the np QCD thermodynamics description at temperatures above deconfinement within FCM was improved by incorporating the color-magnetic confinement (CMC) \cite{20,19,21,22,23}.
CMC produces the effective gluon mass, the Debye screening mass \(m_D\), that linearly grows with temperature above deconfinement. 

With this improvement, a solution to the famous Linde problem was suggested in \cite{19}. 
A compatible with lattice data SU(3) thermodynamics description was constructed in \cite{20,21,22}.
An improved EoS for \(n_f=2+1\) QCD at nonzero baryon density   was derived in \cite{23}.
In the zero density limit, the results           \cite{20,19,21,22,23} are in good agreement with lattice data \cite{24},  using the Debye mass  found in  \cite{25}.
At nonzero density, the absence of the critical point in \((\mu,~T)\)-plane was demonstrated, which is in agreement with lattice data analysis.
In addition, the sound speed in QGP at zero and nonzero density was determined in \cite{26} and \cite{27}, respectively. For a summary of FCM results at  nonzero  density see \cite{Khaisim}.

The topic of the  np dynamics behind the Polyakov loop was suggested in \cite{28}.

The important subject of  thermodynamics in magnetic field was studied in detail in lattice and analytic calculations in \cite{an1,an2,an3,an4,an5,an6, ant1,ant2,thLatt2} at zero baryonic density.

In the present paper, we continue studying QCD thermodynamics in a uniform  magnetic field within FCM \cite{31,32,33,34} applying our methods to nonzero  density.
We start with the quark contribution to the dense deconfined QCD free energy with np interaction incorporated by means of CMC and Polyakov loops.
Then we replace the quark energy with the relativistic charge energy in a m.f to obtain an expression for the free energy in a m.f.
Finally, we derive a EoS for the finite baryon density QGP in a m.f., or expression for the QGP pressure, from the free energy expression;
we define the QGP magnetic susceptibility with the quark contribution to the QGP pressure.
We check our results in a strong m.f.~at the zero density limit by numerical comparison with lattice data.

The paper is organized as follows. 
In section \ref{sect_press} we outline a dense QGP EoS derivation in the absence of a m.f. 
In section \ref{sect_magpress} we generalize the EoS to the case of a nonzero m.f.
In section \ref{sect_susc} we calculate the QGP magnetic susceptibility.
In section \ref{sect_num} we present a numerical evaluation of our analytical results. 
Section \ref{sect_discuss} contains conclusion and discussion.

\section{Quark pressure with Polyakov loops and CMC}\label{sect_press}

We start with a \(f\)-flavored quark (the index \(f=u,d,s...\) is suppressed unless otherwise stated) free energy in background color and electromagnetic fields $A_\mu (x)$ and $A^{(e)}_\mu (x)$ (for the gluon contribution to the QGP pressure, see \cite{20})

 \be \frac{1}{T} F_q (A,A^{(e)} ) = - \frac12 \tr \int^\infty_0 \xi (s)
 \frac{ds}{s} d^4 x\overline{(Dz)}^w_{xx} e^{-K-sm_q^2} \lan W_\sigma
 (C[z])\ran, \label{1}\ee
where $K=\frac14 \int^s_0 \left( \frac{dz_\mu}{d\tau} \right)^2 d \tau$ is the kinetic kernel (we utilized the Feynman-Fock-Schwinger proper time formalism), $m_q$ is the current quark mass, and 

\be W_\sigma (C[z]) = P_F P_A \exp \left(ig \int_{C[z]} A_\mu dz_\mu + ie \int_{C[z]}
A_\mu^{(e)} dz_\mu\right) \exp \int^s_0 \left(g\sigma_{\mu\nu} F_{\mu\nu} +
e\sigma_{\mu\nu} F_{\mu\nu}^{(e)}\right)  d\tau\label{2}\ee 
is a path- and surface-ordered Wilson loop;  $\langle ... \rangle$ denotes the averaging over the stochastic color background field. 
The integration measure implies periodic boundary conditions (\(\beta=T^{-1}\) is the inverse temperature) on a contour $C[z]$ (for details, see \cite{20,23})
\be
\overline{(Dz)}^w_{xy} = \prod^n_{m=1} \frac{d^4\Delta
z_k(m)}{(4\pi\varepsilon)^2}\sum_{n=0,\pm 1,\pm 2} (-)^n \frac{d^4p}{(2\pi)^4}
e^{ip_\mu(\sum\Delta z_\mu(m) - (x-y) - n \beta \delta_{\mu 4})}.\label{3}\ee

We write the series over the Matsubara frequencies in the form 
  \be
P_q = 2 N_c\int^\infty_0  \frac{ds}{s} e^{-m^2_q s} \sum^\infty_{n=1}
(-)^{n+1}[S^{(n)} (s) + S^{(-n)} (s)],\label{4}\ee 
where
\be S^{(n)} (s) = \int
(\overline{Dz} )^w_{on} e^{-K} \frac{1}{N_c} \tr  W_\sigma (C[z]).\label{5}\ee
The T-independent \(n=0\) term was subtracted. 

At this point, we introduce the main np thermodynamics ingredients --- CMC and the Polyakov loop \(L\).
Within the FCM assumptions, color-electric and color-magnetic stochastic background fields are statistically independent. 
Hence, the Polyakov loop and the spatial 3d projection of the quark Green's function \(S_3(s)\), 
which is subjected to CMC, factorize
  \be S^{(n)}(s) =  \int (Dz_4)^w_{on}  e^{-K} L^nS_3(s) = \frac{e^{\frac{-n^2\beta^2}{4s}}L^n}{\sqrt{4\pi s}} S_3(s) .\label{6}\ee

We introduced the quark density by replacing the free energy in \eqref{1} with the thermodynamic potential \(\Omega=F-\mu N\) 
(\(\mu\) is the f-flavored quark chemical potential).
The pressure definition is $P = -\left( \frac{\partial \Omega}{\partial V} \right)_{T,\mu}$, so 
\be  P_q = \frac{ 4N_c}{\sqrt{4\pi}} \int^\infty_0 \frac{ds}{s^{3/2}} e^{- m_q^2s} S_3 (s) \sum_{n=1,2,...} (-)^{n+1} e^{-\frac{n^2}{4T^2s}}\cosh\left(\frac{\mu n}{T}\right) L^n,
 \label{7}\ee

We may extract the quark Polyakov loop $L=\exp \left(-\frac{V_1 (\infty, T)}{2T} \right)$ from lattice simulations (in which case, we may have to change its normalization) or calculate it within FCM \cite{28} (in which case, we still indirectly use lattice data on deconfinement dynamics). 

Following \cite{20,23}, we utilize the approximation
\be S_3(s) \simeq \frac{1}{(4\pi s)^{3/2}} e^{- \frac{m^2_Ds}{4}}, ~~ m^2_D = c_D^2\sigma_s (T),\label{9}\ee
where \(c_D\sim 2\) \cite{25} is a free parameter in our framework, and the color-magnetic string tension $\sigma_s(T) \sim g^4 T^2$ calculated in \cite{25} defines magnetic Debye screening mass $m_D$. 

Therefore, the EoS is
\be \frac{P_q}{T^4} = \frac{N_c}{4\pi^2} \sum^\infty_{n=1} \frac{(-)^{n+1}}{n^4} L^n\cosh \left(\frac{\mu n}{T}\right) \Phi_n (T), \label{10}\ee
where $K_2$ is the Macdonald function
\be \Phi_n (T) = \frac{8n^2\bar M^2}{T^2} K_2 \left( \frac{\bar M n }{T} \right), ~~ \bar M = \sqrt{m^2_q + \frac{m^2_D}{4}}.\label{11}\ee

To reveal the EoS analytical structure in the complex \(\mu\)-plane, we cast the series to the integral form following \cite{23}
\be \frac{P_q}{T^4} = \frac{N_c}{\pi^2} (\xi_+ + \xi_-),\label{13}\ee
\begin{align}
    \xi_\pm &= \sum_{n=1}^\infty \frac{(-)^{n+1}}{n^2} L^n e^{\pm \mu n\beta} \left(\frac{\bar M}{T}\right)^2 K_2 \left( \frac{\bar M n}{T}\right)\label{12}\\
            &= \frac{1}{12} \left( \frac{\bar M}{T} \right)^4 \int^\infty_0 \frac{u^4du}{\sqrt{1+u^2}}\frac{1}{1+\exp \left( \frac{\bar M}{T} \sqrt{1+u^2} + \frac{V_1}{2T} \mp \frac{\mu}{T} \right)}.\label{14}
\end{align}
Cuts in the complex plane immediately follow from \eqref{14}: \(\frac{|\Re\mu|}{T}\ge\frac{\bar M}{T}+\frac{V_1}{2T}\), \(\frac{\Im\mu}{T}=\pi(2n+1)\).
  
In what follows, we also use another form \cite{33} of EoS \eqref{10} 
\be P_q = \frac{2 N_c}{\sqrt{\pi}} \int \frac{ d^3p}{(2\pi)^3} \sum^\infty_{n=1} (-)^{n+1} \sqrt{\frac{2}{\beta n}}  
    L^n \cosh\left(\frac{\mu n}{T}\right)  \int^\infty_0 \frac{d\omega}{\sqrt{\omega}} 
    e^{-\left( \frac{\bar M^2+\vep^2}{2\omega} + \frac{\omega}{2} \right) n\beta}.\label{15}
\ee

\section{Quark pressure in a magnetic field}\label{sect_magpress}

Let us adjust \eqref{15} to the case of a uniform m.f.~directed along the z-axis.

The m.f.~alters quarks motion in the transverse plane.
Transverse motion appears in \eqref{15} through the energy spectra and the phase space.
To make up for the m.f., we introduce the Landau energy levels and modify the phase space following the standard prescription \cite{35}

\be \varepsilon\to\varepsilon^\sigma_{n_\bot}(p_z)= \sqrt{ p^2_z+ \bar M^2 + |e_q B | (2n_\bot +1 - \bar \sigma)},~~ \bar \sigma= \frac{e_q}{|e_q|} \sigma .\label{16}\ee
\be 2\int\frac{V_3d^3\vep}{(2\pi)^3}\to \sum_{\sigma=\pm1}~\sum_{n_\bot=0}^{\infty}\int\frac{dp_z}{2\pi} \frac{|e_q B|}{2\pi} V_3.\label{18}\ee

The standard pressure definition may require an elaboration in case of a non-trivial phase space.
However, the quark pressure along the m.f.~is well-defined in the present case
\be P_z = -\left(\frac{\partial \Omega}{\partial V_z} \right)_{T,\mu},\quad dV_z=S_{\perp}dz.\label{pres} \ee
See Section 5 for the discussion of pressure anisotropy in magnetized QGP.
In what follows, we suppress the subscript, and write \(P\equiv P_z\).

With this reservation, we obtain
\begin{gather}
    P_q(B) = \sum_{n_\bot, \sigma} N_c   T \frac{|e_qB|}{2\pi}
    \frac{\chi(\mu)+\chi(-\mu)}{8},\label{19}\\
    \chi(\mu)=\int \frac{dp_z}{2\pi} \ln
    \left(1+\exp \left( \frac{\mu -V_1/2-\varepsilon^\sigma_{n_\bot}(p_z)}{T}\right)
    \right).  \label{20}
\end{gather}
The integration yields (\(\varepsilon^\sigma_{n_\bot}\equiv\varepsilon^\sigma_{n_\bot}(0)\))
\be P_q (B,T) = \frac{N_c  |e_qB|T}{\pi^2} \sum_{n_\bot, \sigma} \sum^\infty_{n=1} \frac{(-)^{n+1}}{n}  
L^n \cosh \left( \frac{\mu n}{T} \right)  \varepsilon^\sigma_{n_\bot} K_1 \left( \frac{n
\varepsilon^\sigma_{n_\bot}}{T}\right),\label{21}\ee 
and the summation of Landau levels  over $n_\bot$  
yields the expression for f-flavored quark pressure at finite density in a m.f.~(along the m.f.)
\begin{align}
    P_q (B) = &\frac{N_c e_q BT}{\pi^2} \sum^\infty_{n=1} \frac{(-)^{n+1}}{n}L^n \cosh\left(\frac{\mu n}{T}\right) 
    \left(\bar M K_1 \left( \frac{n\bar M}{T}\right)+\right.\nonumber\\
    &+\left.\frac{2T}{n} \frac{e_qB+\bar M^2}{e_qB} K_2 \left( \frac{n}{T} \sqrt{e_q B + \bar M^2}\right) 
    - \frac{ne_qB}{12T} K_0\left( \frac{n}{T} \sqrt{e_q B + \bar M^2}\right)\right).\label{31}
\end{align}

The accuracy of the $n_\bot$  summation in (\ref{21}) exemplified in (\ref{31}) is checked below in section V and Fig.1.

The total quark pressure is the sum over the flavors 
\be P^{(tot)}_q (B,T) = \sum_f P_q^{(f)}(B,T).\label{27}\ee

As in the previous section, we cast the series to the integral form (see the Appendix of \cite{33} for details)
\be \sum_{n_\bot, \sigma}  \chi(\mu) = (I_1+ I_2+I_3),\label{23}\ee
where 
\begin{gather}
    I_1 = \frac{1}{\pi T}\int^\infty_0 \frac{ p_z d p_z}{1+\exp \left( \frac{\sqrt{p^2_i+\bar M^2}-\mu+V_1/2}{T} \right)},\label{24}\\
    I_2 = \frac{1}{\pi T}\int^\infty_0 \frac{ p_z^2 d p_z} {|e_q B|} \int^\infty_0 \frac{ d\lambda}{ \sqrt{p^2_z+\bar M^2 +|e_q B|+\lambda}}
    ~\frac{1}{ 1+\exp \left( \frac{\sqrt{p^2_z+\bar M^2 +|e_q B|+\lambda}-\mu+V_1/2}{T}\right)}, \label{25}\\
    I_3 = -\frac{|e_q B|}{24\pi T }  \int^\infty_{-\infty}   \frac{ dp_z}{ \sqrt{p^2_z+\bar M^2 +|e_q B|}}
    ~\frac{1}{ 1+\exp \left( \frac{\sqrt{p^2_z+\bar M^2 +|e_q B|} - \mu+V_1/2}{T}\right)}. \label{26}
\end{gather}
The integral form of \eqref{31} is
\be P_q (B,T) =N_c T \frac{|e_qB|}{2\pi^2}  \sum^3_{i=1} (I_i(\mu) + I_i(-\mu)).\label{28}\ee  

Finally, we probe the result \eqref{28} with the \(B\to 0\) limit.
$I_1(\mu)$ does not depend on $B$, $I_3(\mu, B)$ is $O(|B|)$, and $I_2(\mu, B)$ is $O(|B|^{-1})$, so \(B\) drops out
\be P_q = \frac{N_c T^4}{12\pi^2} \left[ \varphi\left( \frac{\mu - \frac{V_1}{2}}{T}, \frac{\bar M}{T}\right) 
    + \varphi\left( -\frac{\mu +\frac{V_1}{2}}{T}, \frac{\bar M}{T} \right)\right]\label{29}\ee
\be \varphi(a, \nu) = \int^\infty_0 \frac{z^4 dz}{\sqrt{z^2+\nu^2}} \frac{1}{\exp (\sqrt{z^2+\nu^2} -a)+1}.\label{30}\ee
As expected, the expression coincides with the corresponding result \eqref{13} of section \ref{sect_press}.

\section{QGP magnetic susceptibility at nonzero baryon density}\label{sect_susc}
  
We use the following definition of magnetic susceptibility \cite{an3} of a thermal medium
\be  P(B,T) - P(0,T) = \frac{\hat\chi}{2}(eB)^2 + O((eB)^4). \label{34}\ee

The gluon part of the QGP free energy is sensitive to a m.f.~via the quark loops.
The corresponding contribution to \(\hat\chi\) is suppressed at least as \(\alpha_s^2\).
In the present paper, we neglect this contribution
\be \hat\chi\simeq\hat\chi_q^{(tot)}.\ee
In a sense, this approximation is similar to the staggered fermions approach.

We also introduce the f-flavored quark susceptibilities \(\hat\chi_q^{(f)}\) 
\be  P_q^{(f)}(B,T) - P_q^{(f)}(0,T) = \frac{\hat \chi_q^{(f)}}{2}(e_q^{(f)}B)^2 + O((eB)^4), \label{34*}\ee
which are related to the total quark susceptibility as
\begin{equation}
    \hat\chi_q^{(tot)}= \sum_f\hat\chi_q^{(f)}(e_q^{(f)}/e)^2.\label{34**}
\end{equation}
We extract the susceptibilities from \eqref{31} by expanding it in powers of \(e_qB\)
\be  \hat \chi_q = \frac{   N_c}{3\pi^2} \sum^\infty_{n=1}  (-)^{n+1}
    L^n\cosh \left(\frac{\mu n}{T}\right) K_0 \left( \frac{n \bar M}{T} \right). \label{35}\ee
  
With the integral representation for the Macdonald function
\be K_0 (z) = \frac12 \int^\infty_0 \frac{dx}{x} e^{-\left( \frac{1}{x} + \frac{z^2 x}{4} \right)},\label{36}\ee
the integral form of the susceptibilities was obtained
\be  \hat \chi_q =\frac{   N_c}{3\pi^2}  (I_q(\mu) + I_q(-\mu)), \label{37}\ee
\be I_q (\mu) = \frac12 \int^\infty_0 \frac{dx }{x} \frac{L e^{(\mu/T)} e^{-\left( \frac{1}{x} + \frac{\bar M^2  x}{4T^2} \right)}}{1+ L e^{({\mu}/{T})} e^{-\left( \frac{1}{x} + \frac{\bar M^2  x}{4T^2} \right)}}\label{38}\ee

\section{Numerical analysis of the results}\label{sect_num}

We computed thermodynamic quantities of QGP with u-, d-, s-quarks by means of corresponding series partial summation.
The np inputs $V_1(\infty,~T)$ and \(m_D(T)\) are identical to these of \cite{Khaisim}.
They were adjusted in such a way that QGP pressure \eqref{10} at zero m.f.~and baryon density matches the lattice data \cite{qlatt2,latt3}.
The Debye mass is defined by the spatial string tension value \(m_D= c_D \sqrt{\sigma_s(T)}\); for \(\sigma_s(T)\) see (45) in \cite{25}.
The parameter \(c_D=1.6\) is close to the corresponding result of \cite{25}.

The major source of systematic error is $V_1(\infty,~T)$ uncertainty.
Within the parameters range, we estimate the total error of this section numerical results to be at most \(15\%\) (see section \ref{sect_discuss} for details).

\begin{figure}
    \center{\includegraphics[width=0.7\linewidth]{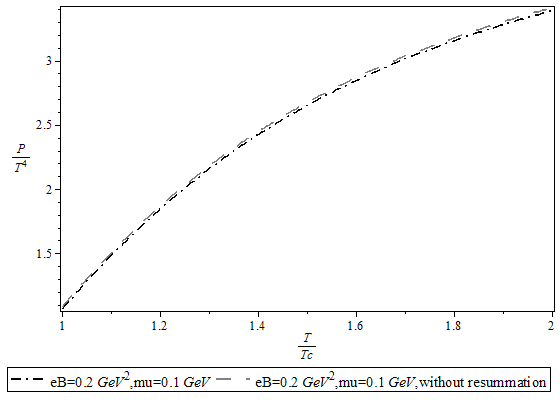}}
    \caption{The  pressure $P/T^4$ from  \eqref{21}, the dash-dotted line,  compared with the integral  form, Eq. (\ref{31}) the dashed line.}
    \label{fig_summ}
\end{figure}

Let us start with an accuracy check of the approximate Landau levels summation procedure.
The Fig.~\ref{fig_summ} demonstrates the difference between the result of \eqref{21} based on the partial sum  and   the integral form in \eqref{31}.
We find the difference to be negligible in the considered parameter range.

\begin{figure}
    \center{\includegraphics[width=0.7\linewidth]{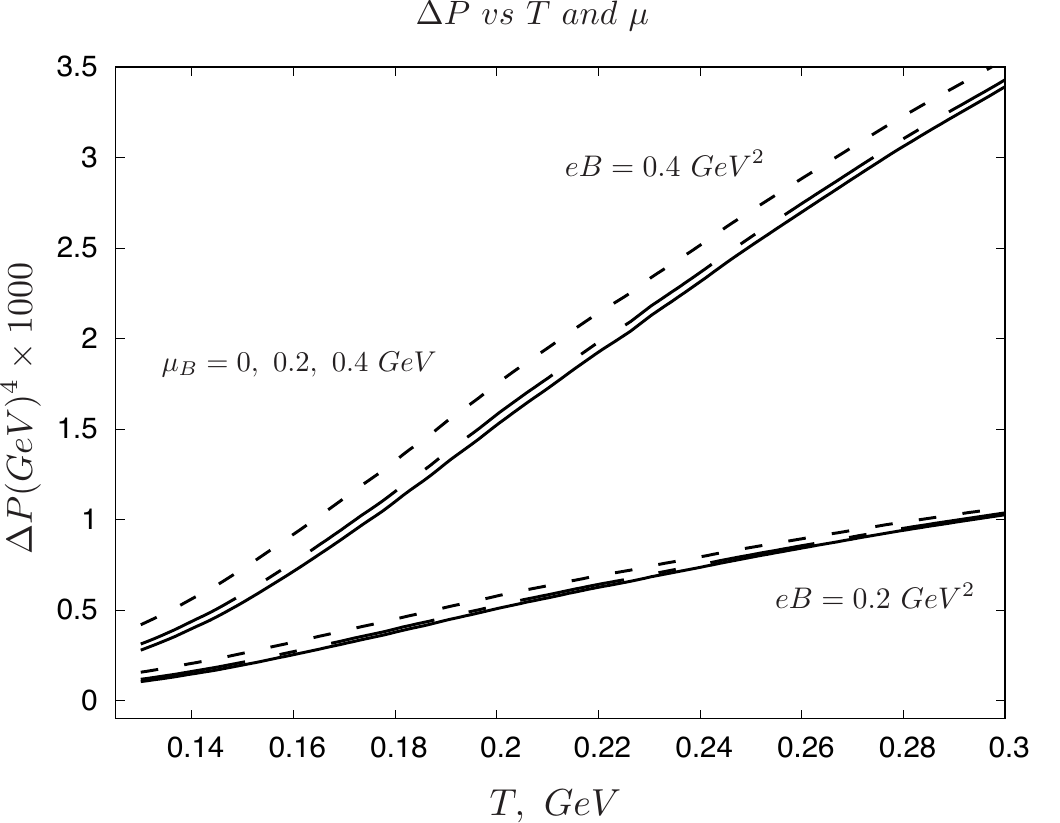}}
    \caption{The magnetic shift $\Delta P = P(eB,\mu) - P(eB = 0,\mu)$ for dense QGP with u,~d,~s quark flavors. 
            For a given m.f., each trajectory demonstrates triple splitting for different $\mu_B$. 
            the Solid lines are for $\mu_B = 0$, the dash-dotted --- for $\mu_B = 0.2~GeV$, and the dashed --- for $\mu_B = 0.4~GeV$.}
    \label{fig1}
\end{figure}

The set of trajectories for the pressure magnetic shift $\Delta P = P(eB,~\mu_B) - P(0,~\mu_B)$, computed with \eqref{31}, is presented in Fig.~\ref{fig1} 
(baryon and quark chemical potentials are related as $\mu_f = \frac{1}{3}\mu_B$, $f = u,d,s$). 
The results suggest strong paramagnetism of QGP, in agreement with the lattice study \cite{an2}.

With the expressions for gluon pressure from \cite{31} and for quark pressure \eqref{31}, we computed a special combination \(\Delta\) for dense QGP in an external m.f.
\be
    \Delta = \frac{\epsilon-3P_z+\mu n}{T^4}.
\ee
If pressure is considered as isotropic,   \(\Delta\) is the trace anomaly.

\begin{figure}
    \includegraphics[width=0.7\linewidth]{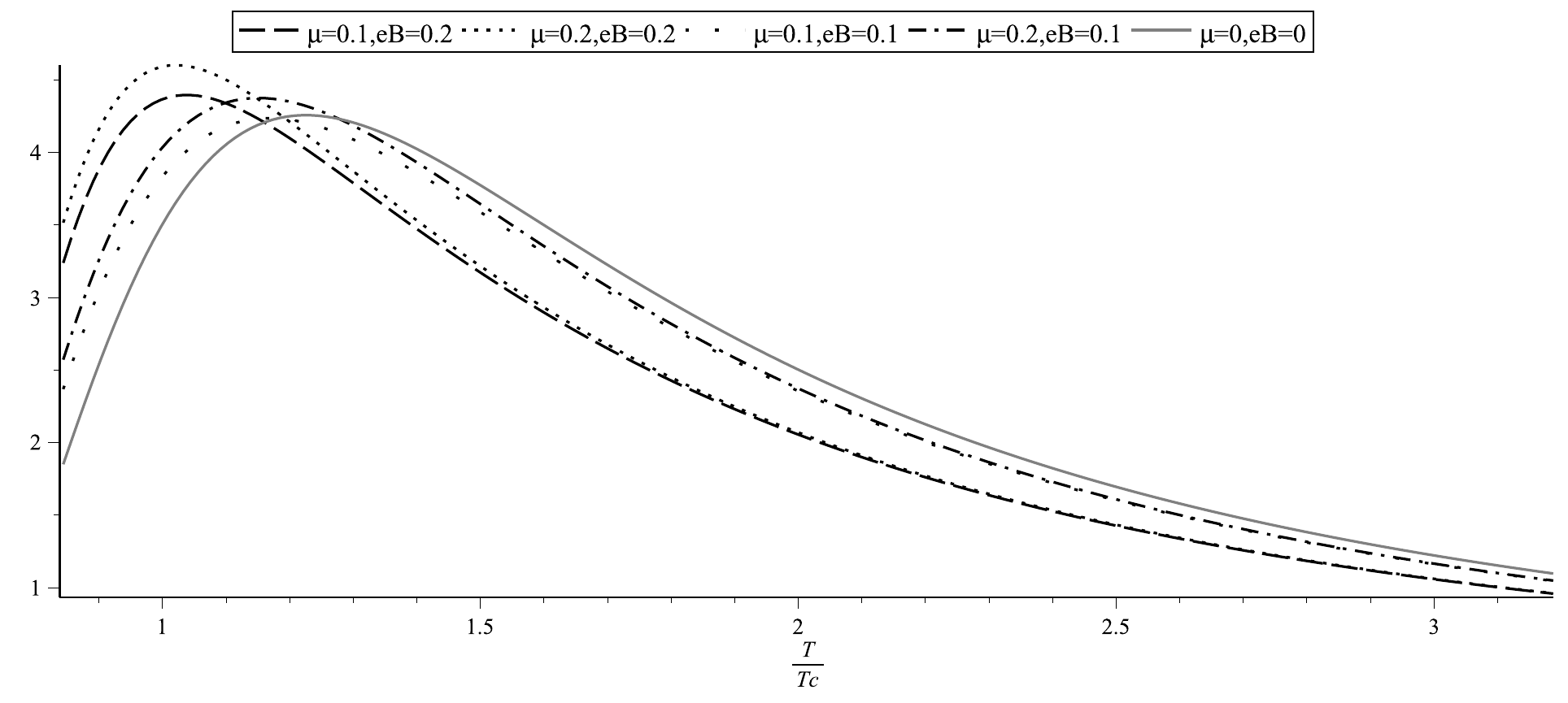}
    \caption{The combination \(\Delta\) for QGP with u,~d,~s quark flavors as a function of \(T\) at various values of \(\mu_B[GeV]\) and \(eB[GeV^2]\).}
    \label{fig_anomaly}
\end{figure}

The resulting curves in Fig.~\ref{fig_anomaly} demonstrate all the qualitative features (the maximum position and width dependence on m.f., the curves intersections) of the same quantity computed with the lattice simulation \cite{an6} (see Fig.~12 of \cite{an6}) in the given temperature range. 

\section{Conclusion and Discussion}\label{sect_discuss}
In the present paper, pressure of QGP at nonzero baryon density in an external uniform m.f.~was calculated analytically for the first time.

Within the parameters range (\(eB<0.5~GeV^2,~\mu_B<0.5~GeV\)), we neglect the vacuum properties alternation.
That is, we neglected the Polyakov loop and the Debye mass dependencies on the baryon density and the m.f. 
According to the recent lattice data, the approximations \(L(T,\mu_B,B)\to L(T),~m_D(T,B)\to m_D(T),~m_D(T,\mu)\to m_D(T)\) yield up to 7\% 
\cite{09547}, 5\% \cite{00842}, and 2\% \cite{09461} errors, respectively.
We consider the errors as independent, and estimate  the total EoS \eqref{31} error in ``the worst'' parameters region (low temperature, high m.f.~and density) as 15\%.  
At larger densities, the Polyakov loop dependence on the chemical potential is to be accounted for.

The Polyakov loops interaction and the perturbative corrections were also neglected.
The non-perturbative numerical input data were set as to reproduce the correct QGP pressure values at the zero m.f.~and baryon density limit.

With the numerical analysis of our results, we observed QGP strong paramagnetism, which was previously found within lattice simulations.
The possible pressure anisotropy, on the other hand, is a matter of discussion.  

QGP pressure anisotropy is observed in lattice simulations with a certain setup.
Basic methods for pressure calculation on lattice are \cite{an1} 
\begin{enumerate}
\item $B=const$ method --- the variation of energy in a constant external m.f.~(the QGP is assumed to be in thermal equilibrium);
\item $\Phi=const$ method, in which the total magnetic flux $\Phi$ through the lattice is fixed. 
\end{enumerate}
With the first method, pressure is isotropic, while with the second method 
\begin{equation}
P_x = P_y = P_z - M \cdot B, \label{39}	
\end{equation}
where $M = -\frac{1}{V}\frac{\partial F}{\partial B}$ is the QGP magnetization.

The ``anisotropic'' method is more subtle for real physical systems. 
It requires a careful analysis of the energy exchange between longitudinal and transverse motion of quarks and the corresponding relaxation time. 
We leave these details to a future publication. 

The direct influence of a m.f.~on the field correlators was studied on the lattice in  \cite{07012}. 
Since the effect is at most 10\% at \(eB<0.5~GeV^2\), we had assumed it to be negligible for our purposes. 
The more exact accuracy analysis with the Polyakov loop and the screening mass variation justified this assumption. 

\section{Acknowledgments}
The authors are grateful to B.O.~Kerbikov and M.S.~Lukashov for useful discussions, and to N.P.~Igumnova for technical help in the manuscript preparation.
This work was supported by the Russian Science Foundation Grant number 16-12-10414.

\end{document}